\documentclass{iopconfser}

\usepackage{graphicx}
\usepackage{textgreek}
\usepackage{subfigure}
\usepackage{amsmath}
\usepackage{float}
\usepackage{multirow}
\usepackage[style=numeric,sorting=none]{biblatex}
\usepackage{color,soul}

\DeclareMathOperator{\cov}{cov}

\begin{document}

\title{Analytically Optimising Muon Diffusion Experiments with Fisher information}

\author{Alex Sampson$^{1,2, *}$, Peter J. Baker$^{1}$, Lucas Wilkins$^{1}$, 
    John M. Wilkinson$^{1, \dagger}$}

\affil{$^1$ISIS Neutron and Muon Source, STFC Rutherford Appleton Laboratory, Harwell Campus, Oxfordshire, OX11 0QX, United Kingdom}
\affil{$^2$School of Computer Science, The University of Sheffield, Regent Court,
211 Portobello, Sheffield, S1 4DP, United Kingdom}

\email{$^*$asampson1@sheffield.ac.uk, $^\dagger$john.wilkinson@stfc.ac.uk}

\begin{abstract}
One of the key challenges in performing muon experiments is knowing which temperatures and applied fields to measure at, and how many muon decays should be measured at each temperature/field combination to get the most useful dataset. We have developed a technique using Fisher information which, for a given muon asymmetry function, can analytically calculate the number of muon decays required to obtain a given error on the parameters of the asymmetry model. Here, we report on the results of our project, in particular applying our methodology to the problem of knowing the best choice of applied longitudinal fields for ionic diffusion experiments.
\end{abstract}

\section{Introduction}

One of the greatest challenges in undertaking experiments at large-scale 
facilities is the fundamental economic problem that 
beamtime is scarce and the number of ways to measure a sample is very large.
To optimise the scientific output of the facility, the scientific aims of the 
experiments need to be enabled by collecting the most data of sufficient statistical 
significance in the available time on the instruments.
In many $\mu$SR experiments, the experimenter needs to select the 
temperatures and fields that should be applied to the sample to make the best 
use of their experimental time, but these parameters are often 
decided using only experience and `rules of thumb'. Recent advances in 
neutron scattering show that it is possible to use the Fisher information 
to optimise such experiments \cite{durant2022}. 
Here, we show that Fisher information can also be used to optimise muon 
experiments, in particular optimising the chosen field strengths when 
applying longitudinal fields for ionic diffusion experiments 
\cite{McClelland2020} and showing that such an optimisation leads to the 
errors in the  resulting fits being minimised.
This work primarily focuses on \(\mu\)SR experiments on materials with 
either muonic or ionic diffusion resulting in a the muon exhibiting a dynamic 
Kubo-Toyabe polarisation function \cite{KuboToyabe, kadono}, but such 
a technique can be extended to optimise experiments involving materials 
that exhibit any polarisation function.

%


\section{Analytical optimisation method}
\label{sec:analytical_opt}

\subsection{Background}

Fisher information is a quickly obtainable, informational lower bound 
on the errors of fitted parameters~\cite{Rao, Cramer}.
As it is an analytical result, it allows predictions to be made about the 
expected errors with relatively little computational expense. 
Although there are other approaches that address the same core problem
(perhaps more robustly), they require significantly more 
computational effort (as shown in  Ref.~\cite{durant2022}).

In the discrete case, applicable to count statistics, the Fisher information 
can be defined as the negative expectation of the Hessian of the log likelihood,

\begin{equation}
\mathcal{I}_{ij}(\xi) = E\left[\frac{\partial^2}{\partial \xi_i \partial \xi_j} \log p(x; \xi)\right] = - \sum_X p(x; {\xi}) \frac{\partial^2}{\partial \xi_i \partial \xi_j}                     \log p(x; {\xi}) ,
\end{equation}
where $\mathcal{I}_{ij}(\xi)$ represents the \((i,j)\)-th component of the Fisher information matrix with respect to the parameter vector \(\xi\), \(x\) represents a possible observation drawn from the sample space,
and \(p(x; \xi)\) represents the probability distribution of \(x\) parameterised by \(\xi\).

When considering many common distributions with their standard parametrisations, the Fisher information and the covariance of random variables (i.e. of $X$, not the
parameter estimates $\hat{\xi}$) often coincide. For example, the Fisher information for a Poisson distributed variable $X\sim\text{Poi}(\lambda)$ is $\lambda^{-1}$ and the variance of $X$ is $\lambda$. Moreover, in the majority of cases, calculations involving the Fisher information are approximated well by those of error propagation, even though the underlying mathematical machinery is quite different. 

Indeed, whilst there are many superficial similarities, the Fisher information is not the same as parameter errors, but is 
linked to them by a central result in statistics: the Cram\'er-Rao bound ~\cite{Rao, Cramer}.
This states that the covariance ($\cov\hat{\xi}$) of an estimate ($\hat{\xi}$) of parameters ($\xi$) is bounded by Fisher information in the sense that the matrix valued quantity 
$\mathcal{I}(\xi) - \cov(\hat{\xi} - \xi)$
is positive semi-definite. For unbiased estimates (i.e. where $E(\hat{\xi}) = \xi$), this relationship is often summarised with the notation
\begin{equation}
\mathcal{I}(\xi)^{-1} \preceq \cov(\hat{\xi}).
\label{eq:cramer-rao}
\end{equation}
The $\preceq$ symbol changes between different authors, c.f.~\cite{Rao}, but is generally chosen to convey a multidimensional generalisation of an inequality (see ~\cite{durant2022} for a graphical interpretation.)
Once we recognise that we only have a bound on what we expect in practice, it becomes an empirical question as to whether these bounds are useful -- one which we answer in the affirmative in the subsequent sections of this paper.

Often the relationship between the Fisher information and estimate covariance is described by the efficiency, $e$, so
\begin{equation}
    \mathcal{I}(\xi)^{-1} = e(\hat{\xi}) \cov(\hat{\xi}).
    \label{eq:fullcr}
\end{equation}
Generally speaking, numerical estimates of parameters will not be perfectly efficient, using all the information available in the data (i.e. achieve equality in the Cram\'er-Rao bound). In the case of predicting the number of counts needed to get a particular error, we would
like $e$ to be simple, preferably constant. But for optimising experimental conditions, we simply require that there be an approximately monotonic relationship, where smaller Fisher information corresponds to larger bounds.

\subsection{Calculation}

The probability distribution associated with a general muon asymmetry function 
$A(t)=\frac{N_{\rm F}(t)-\alpha N_{\rm B}(t)}{N_{\rm F}(t)+\alpha N_{\rm B}(t)}$ is
difficult to work with, as it is supported over the quotients, and to our knowledge
no closed form expression for the corresponding Fisher information exists\footnote{Even the unnormalised version of the asymmetry function 
$(N_\text{F} - \alpha N_\text{B})$ when simplified with $\alpha=1$ has no known closed form for the Fisher information, despite it being described by the relatively simple Skellam distribution.}.
However, we can make the approximation that this probability distrubution approaches 
a normal distribution for sufficiently many counts in both the 
forward and backward detectors. This normal distribution has a variance $\sigma_i^2$ 
given at time $t_i$  after implantation, by
\begin{equation}
\sigma_i(A)^2 = \frac{ e^{\frac{t_i}{\tau_{\mu}}}}{2N_0}(1 - A(t_i)^2),
\end{equation}
where \(N_0\) is the number of muon decay events at $t=0$ and 
$\tau_{\mu}=2.2$~\textmu s
is the average lifetime of a muon. The experiment time $t_i$ for 
our case is 
binned into 16~ns intervals and $t_0$ is the first of these bins, 
which is about 32~ns for a pulsed source (corresponding to the end 
of the incoming pulse of muons).

The Fisher information for a normal distribution with a mean-dependent standard deviation $\sigma(\mu)$ has the general form of
\begin{equation}
    \mathcal{I}(\mu) = 
       \frac{1}{\sigma(\mu)^2} + 
       2 \left(\frac{\partial}{\partial\mu}\log \sigma(\mu) \right)^2.
\end{equation}
Note that in the second term, the constant multiplicative term in the variance, $e^{t_i/\tau_\mu}/2N_0$, becomes additive through the action of the logarithm and
thus becomes zero after differentiation. The first term grows with the number of counts, but the second term does not, so for a sufficiently large number of counts (which we have already assumed to get to this point) only the first term matters.

We now consider the joint probability distribution of the asymmetries for each of the time
bins $t_i$. The probability distribution in each bin is governed by the asymmetry, and 
has Fisher information as described above. As each bin is statistically independent,
the Fisher information for the joint distribution is simply
\begin{equation} 
\mathcal{I}_{ij}(A) = \delta_{ij} \frac{2N_0}{1 - A(t_i)^2} e^{- {\frac{t_i}{\tau_{\mu}}}}.
\end{equation}

To get the Fisher information associated with a particular model of asymmetry over time,
we take a projection of this high-dimensional asymmetry space parameterised by asymmetry in a subspace parameterised by out model parameters, $\xi$. 
As the Fisher information is a metric tensor this reparameterisation can be done according standard tensor coordinate transforms

\begin{equation}
\mathcal{I}_{ab}(\xi) = \sum_{i,j} \frac{\partial A(t_i)}{\partial \xi_a} \frac{\partial A(t_j)}{\partial \xi_b}\mathcal{I}_{ij}(A).
\label{eq:FImatrix}
\end{equation}
For the case of a dynamical Kubo-Toyabe function, 
\( \xi = \{ \nu, \Delta \} \) where \(\nu\) and 
\(\Delta\) represent the fluctuation rate and the second moment of the
field distribution, respectively.

To get an estimate of the minimum number of counts we require to get a prescribed degree 
of certainty we can then apply Eq.~\eqref{eq:cramer-rao} (i.e. calculate 
$\mathcal{I}^{-1}$) and the number of standard deviations according to this lower bound can be used as a measure for the error in the estimates of 
their respective parameters. As $\mathcal{I}$ depends on $N_0$, this can be used 
to analytically calculate the number of muon decay events required to obtain 
a certain error in the parameters of the fit. It should be noted that working out
counting time requires more assumptions about the relationship between the Fisher information and the parameter estimates 
(including full knowledge of the efficiency $e$ in Eq.~\eqref{eq:fullcr}), than working out optimal conditions, which only
require the assumption of monotonicity.

The process outlined above using these equations forms the groundwork of the 
mathematics involved in this paper: for a given set of parameters, and 
an asymmetry function, this technique 
can be used to calculate the minimum error that can be achieved for a given 
number of muon decay events.


\subsection{Application to ionic diffusion experiments}
In $\mu$SR experiments measuring ionic diffusion, it is often advantageous to  
take data at the same temperature multiple times at both zero field (ZF) and with
a series of small applied longitudinal fields (LF), so that the effects of the background
(including the sample holder) and the ionic diffusion can be decoupled. Such a 
method will yield parameter estimates with errors smaller than those that can 
 be achieved by counting events exclusively at any of the selected fields. In other
 words, the form of the asymmetry function we will be optimising is 
\begin{equation}
A({A_0, \Delta, \nu, \Delta_{\rm bg}, c};B,t)=A_0(cG_\mathrm{DKT}(\Delta, \nu ;B,t)
+ (1-c)G_\mathrm{DKT}(\Delta_{\rm bg}, 0; B, t)),
\label{fitfunction}
\end{equation}
where $G_{\rm DKT}$ is the dynamic Kubo-Toyabe function in an applied field 
\cite{Hayano}, 
implemented using the method in Ref.~\cite{allodi2014}. The first of these terms is to
represent the material in question, and the second term represents muons stopped
in the sample holder or cryostat tails (which will usually have a very small 
$\Delta$). Unless specified otherwise, for our simulations we take $A_0=22\%$,
$c=0.65$, and $\Delta_{\rm bg}=0.005$~\textmu s$^{-1}$ as is appropriate for a 
titanium holder. We are using the Kubo-Toyabe function for this study because
it has a well-defined relationship with the applied longitudinal field,
and is relatively easy to calculate compared to other microscopic models 
\cite{Celio}.


\section{Optimising the number of applied longitudinal fields}
\label{sec:threefields}

\begin{figure}
    \centering
    \includegraphics[width=1\textwidth]{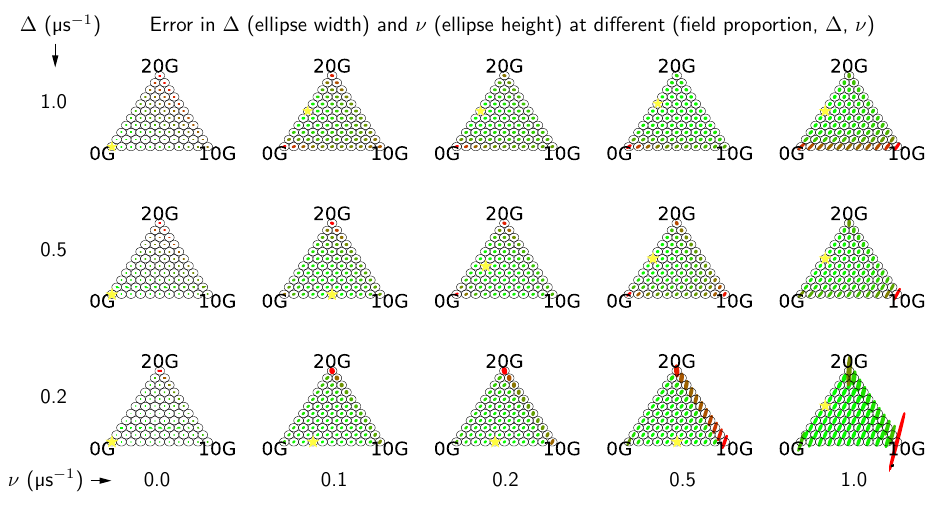}
    \caption{The predicted error in $\Delta$ and $\nu$ (both in 
    \textmu s$^{-1}$) achieved by measuring a series of longitudinally 
    applied fields, visualized as a series of triangular plots where each 
    vertex corresponds to measuring 100\% of the muon statistics at the 
    specified longitudinal field strength. Within each triangle, 
    ellipses are placed at discrete positions (with black circles to 
    represent the centre of these) representing a differing 
    proportion of muon decay statistics measured in the the three field 
    environments. Positions closer to a vertex indicates a larger proportion 
    of measurements taken with the corresponding field. The ellipses 
    within the black circles represent the error covariance matrix as 
    described in the text: the size and colour represent the magnitude
    of the errors (with red ellipses having the largest error, and green
    having the smallest) and the shape of the ellipse represents the 
    structure of the matrix as described in the text. The smallest 
    ellipse in each triangle is marked in blue with a yellow star overlaid
    for clarity.}
    \label{fig:triangle_plots}
\end{figure}

As it is conventional to conduct diffusion experiments by measuring several 
fields at each temperature (one of which is often ZF) \cite{McClelland2020}, here we 
initially optimise experiments with three applied fields. One approach would be
to optimise LF experiments with three applied fields by finding the optimal 
strengths of the two non-zero fields as well as the optimal proportion of muon 
decay events collected at each field, but this introduces five parameters
and is therefore too computationally challenging. We instead run simulations 
(evaluating Eq.~\eqref{eq:FImatrix} with the
asymmetry function in Eq.~\eqref{fitfunction}) to perform 
an optimisation of the proportion of counts conducted at each field strength
$B\in \{0, 10, 20\}$~Gauss with a range of values of $\nu$ and $\Delta$, to 
achieve a user-defined error
(keeping the other parameters in Eq.~\eqref{fitfunction} constant).

The errors in the parameter values for each 
\(\nu\), \(\Delta\)  are depicted in 
Fig.~\ref{fig:triangle_plots} as a set of covariance ellipses in an arrangement 
akin to a ternary plot. This visualisation was made by transforming a set of 
points distributed around the circumference of a unit circle into an ellipse using 
a Cholesky transformation. This effectively adjusts the circle to match the spread 
(variances) and orientation (principal directions) of uncertainty in the 
parameters, as represented by the covariance matrix.  These ellipses are 
scaled along each axis by the reciprocal of the acceptable error in the 
corresponding parameter. The size of the 
final ellipse can then be interpreted as a measure of the size of the predicted
error, with larger ellipses indicating a greater uncertainty in the parameters.

As the smallest ellipse is almost always found on a corner or edge of the triangle 
plots and never in the middle, we suggest that it is rarely optimal to measure 
any more than two applied longitudinal fields: a correctly chosen pair of 
fields is almost always the best way to run the experiment\footnote{The extent to 
which this is optimal depends strongly on $\nu$ and $\Delta$,
but the optimal partitioning of the statistics is almost 
always 50\% in ZF and 50\% in one LF.}. For the only 
example we could find
where this was not the case (where $\Delta=0.5$~\textmu s$^{-1}$ and
$\nu=0.2$~\textmu s$^{-1}$), very few counts are required at one 
of the longitudinal fields and the benefit obtained from measuring 
three fields instead of two is marginal. 

Note that for the case $\nu=0.0$~\textmu s$^{-1}$ (i.e. the static limit),
the Kubo-Toyabe at long 
times tends towards a constant $\frac{1}{3}$ value, meaning the only 
source of relaxation will be due to the sample holder which will 
decouple in applied fields much smaller than 10~Gauss: fields of 
this magnitude will only serve to decouple \emph{both} Kubo-Toyabe
curves leading to little benefit, hence the optimum strategy in this case
is to measure at ZF only. Nevertheless, applying a very small 
(much less than 10~Gauss) longitudinal field can
still be beneficial, as we will show in the next section.

\section{Optimising the most effective longitudinal field to reduce 
parameter uncertainties }
\label{sec:twofield}

\begin{figure}
    \centering
    \subfigure[]{
        \includegraphics[width=0.45\textwidth]{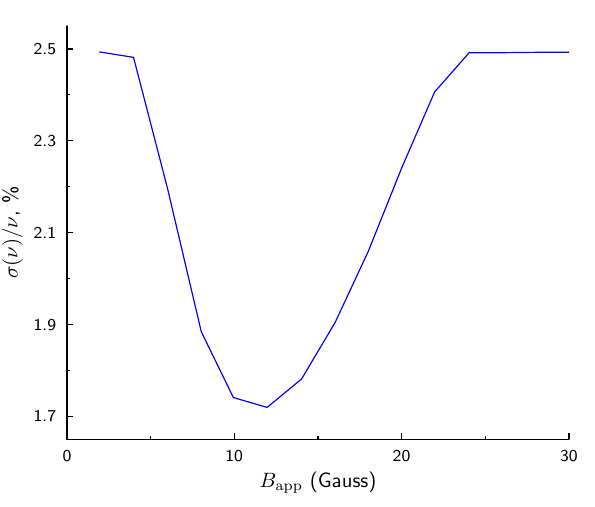}
        \label{fig:error_curve_fi}
    }
    \subfigure[]{
        \includegraphics[width=0.45\textwidth]{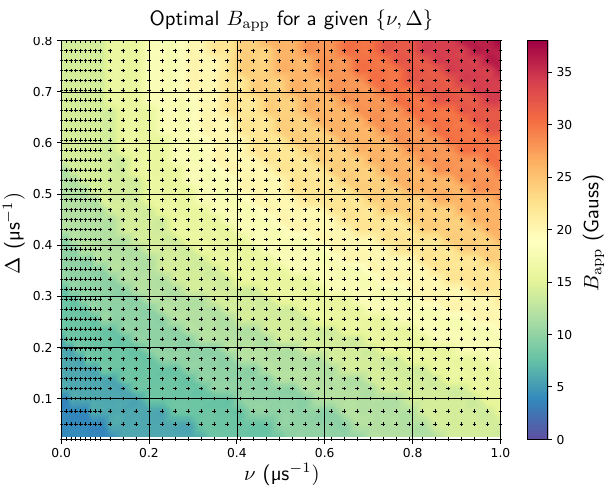}
        \label{fig:heatmap_field}
    }
    \subfigure[]{
        \includegraphics[width=0.45\textwidth]{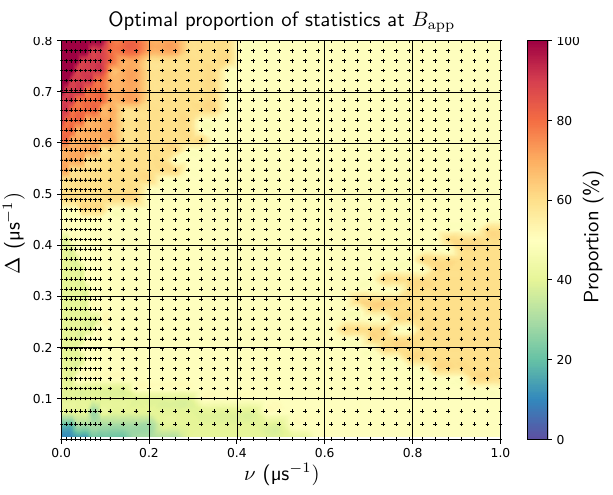}
        \label{fig:heatmap_proportion}
    }
    \caption{(a), the relative error in $\nu$ vs applied field strength ($B_{\rm app}$ in the text) plotted for 
    $\nu=0.1$~\textmu s$^{-1}$ and $\Delta=0.37$~\textmu s$^{-1}$)
    (b), heatmaps over a range of \(\nu\) and \(\Delta\) (both in \textmu s$^{-1}$) of
    optimal longitudinal field strength values (in Gauss) and (c) the percentage of the 
    overall muon decay statistics that should be taken at the optimum non-zero applied 
    field.}
    \label{fig:heatmap}
\end{figure}

As we have seen in Sec.~\ref{sec:threefields}, it is usually optimal to 
only perform a diffusion experiment with two fields (ZF and one LF).
We now turn to the problem of optimising the non-zero applied field such that the
errors in the parameters of the Kubo-Toyabe function are minimised. As this is
less computationally intensive, it allows us to explore a full range of possible 
fields for the non-zero field strength.

Using a method akin to Sec.~\ref{sec:threefields}, we calculate the error in $\Delta$ and 
$\nu$ that one would obtain when simultaneously fitting Eq.~\eqref{fitfunction} with 
fields $B\in\{0,B_{\rm app}\}$, and $p$\% of the total events taken whilst applying 
a longitudinal field $B_{\rm app}$. We then select the field $B_{\rm app}$ which minimises the 
error in $\nu$ and $\Delta$ and the corresponding $p$.
For each \(\nu\) and \(\Delta\) combination investigated, 
 we calculate the error in $\nu$ for each applied $B_{\rm app}$, and we take the minimum of each 
 of these as the optimal field strength for the corresponding 
set of parameters. Fig.~\ref{fig:error_curve_fi} shows a representative example of such 
a calculation for one specific \(\nu\) and \(\Delta\) combination. This behaviour is 
to be expected: if $B_{\rm app}$ is very close to zero, then the applied field will make little
difference to the muon asymmetry, but if $B_{\rm app}$ is too large, the muon's Hamiltonian will
be dominated by the Zeeman contribution of the longitudinal field and will overwhelm the 
nuclear fields we are attempting to measure. 

Fig.~\ref{fig:heatmap_field} 
shows the results of our optimisation as a heatmap for a range of values of $\nu$ and $\Delta$,
which shows that the optimal longitudinal field strength $B_{\rm app}$ increases monotonically with 
$\nu$ and $\Delta$, as expected\footnote{It is nevertheless important to note that 
the there is no simple relationship between the optimal field and $\Delta/\nu$.}. Fig~\ref{fig:heatmap_proportion} shows that the optimal proportion of events 
that should be taken in a longitudinal field of $B_{\rm app}$ is around 50\% for the majority of
values of \(\nu\) and \(\Delta\) we investigated.



%
%
%

\subsection{Application to muon diffusion in copper}

\begin{figure}
    \centering
    \subfigure[]{
        \includegraphics[width=0.45\textwidth]{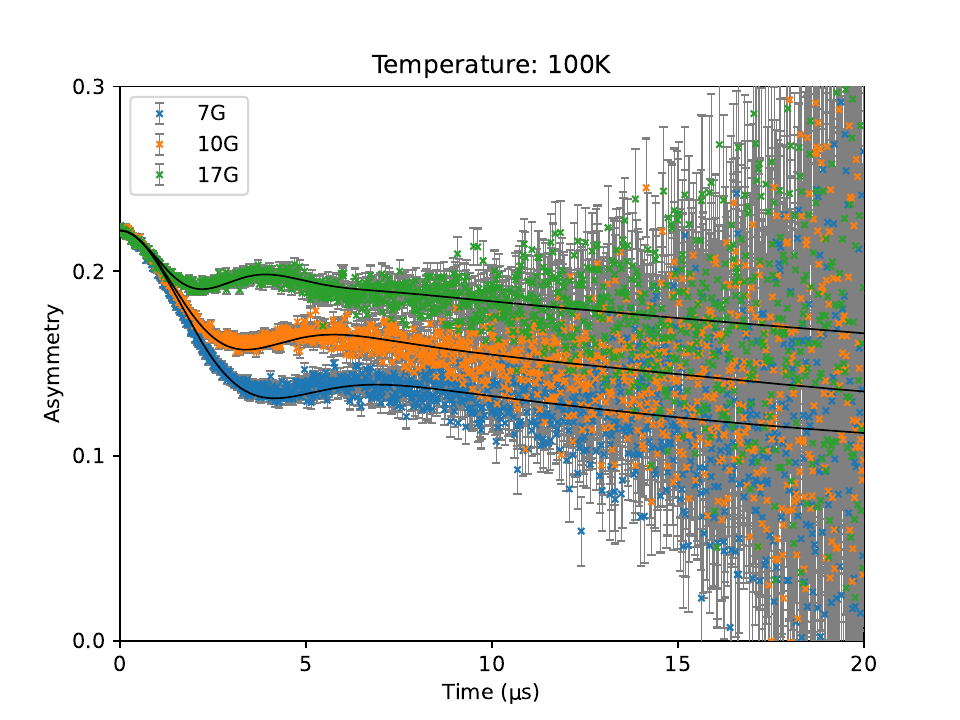}
        \label{fig:100K_data}
    }
    \subfigure[]{
        \includegraphics[width=0.45\textwidth]{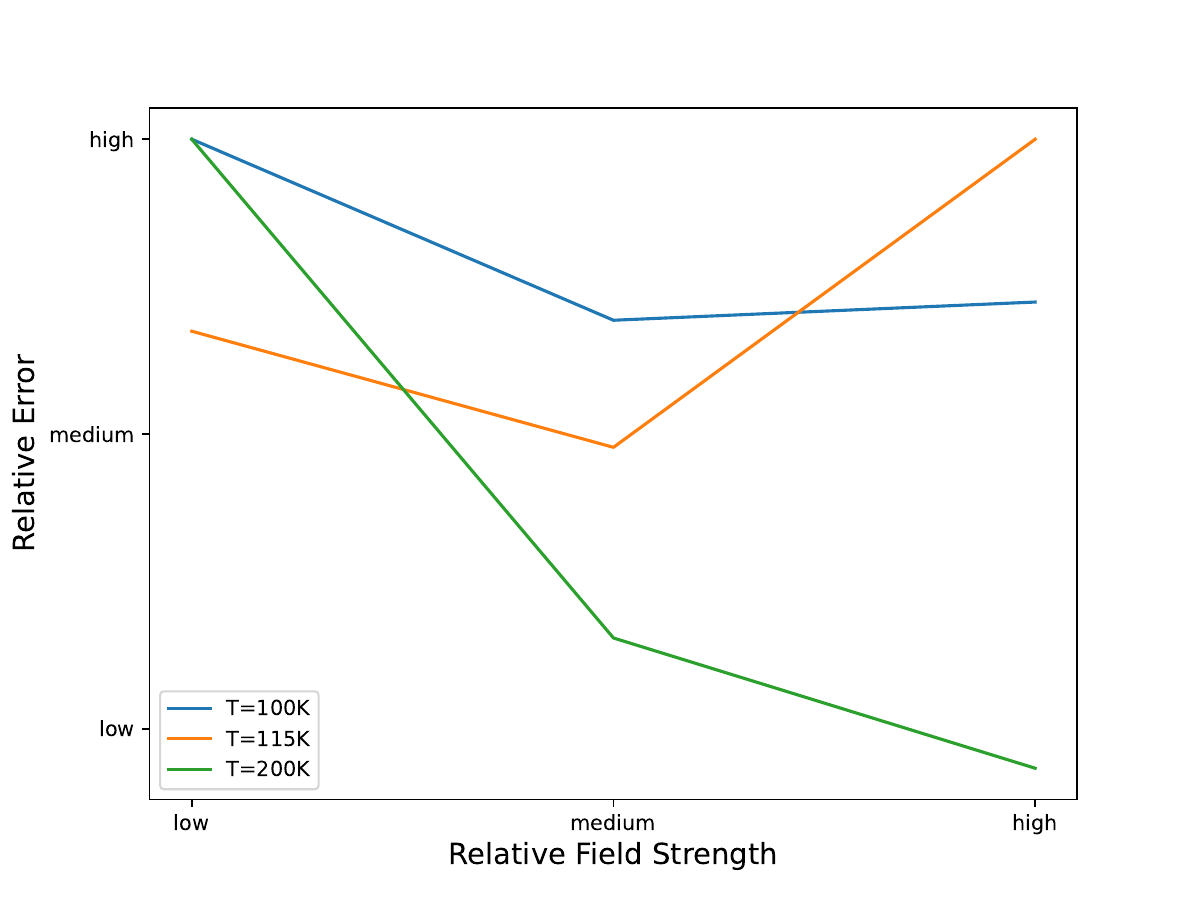}
        \label{fig:scaled_experimental_error_plot}
    }
    \caption{(a), LF data taken at 100~K [the ZF data is omitted for clarity]. (b), scaled 
    plot showing how the relative error in the hopping rate in copper 
    (obtained from the fit to the data), varies with LF strength ($B_{\rm app}$) and temperature.}
    \label{fig:experiment_plot}
\end{figure}

To provide further supporting evidence for the validity of our Fisher information 
approach to planning experiments, we now apply this to an experiment on copper
\cite{data}.
Copper is well known to exhibit muon diffusion~\cite{kadono} at temperatures 
below about 20~K (via quantum tunnelling) and temperatures above about 100~K
(thermal diffusion). We prepared a sample of copper with a titanium plate 
covering a small fraction of the sample (to mimic a typical sample holder
for this type of experiment), and inserted it into the EMU spectrometer 
at ISIS \cite{EMU} in a Variox 
cryostat. We followed the procedure in Sec.~\ref{sec:twofield}, that is 
locating the optimal $B_{\rm app}$ for the expected $\nu$ and $\Delta$ 
(using Ref.~\cite{kadono} as a guide), and measuring at the optimal field, one
field below and another above this. The measured fields listed in Table~\ref{tab:fields}.

\begin{table}
    \centering
    \begin{tabular}{| c | c | c | c |}
        \hline
        \multirow{2}{*}{T (K)}    &   \multicolumn{3}{c|}{$B_{\rm app}$ (Gauss)} \\
        & Low  &  Medium  & High \\
        \hline
        100  & 7.0  & 10.0 & 17.0 \\
        115  & 6.5  & 11.5  & 18.5 \\
        200 & 9.5  & 24.0  & 60.0 \\
        \hline
    \end{tabular}
    \caption{The applied longitudinal fields ($B_{\rm app}$) to the copper sample. The `Medium 
             field' is the optimum field predicted by the method detailed in 
             Section 3.}
    \label{tab:fields}
\end{table}

As part of the data analysis for this experiment, we implemented a bootstrapping 
method \cite{bootstrapping} for taking different samples of the data recorded at 
each of the field strengths and temperatures. In particular, we took 120~million 
muon decay events at each temperature and field in smaller 10 million event runs, 
and simultaneously fit 110-million event combinations of these  (i.e. excluding one 
run each time). For each of these combinations, we simultaneously fit the zero field 
data along with the data at each individual field using Eq.~\eqref{fitfunction}, 
and then taking the standard deviation of the fitting parameters for each of the 
samples (which represents the error at that specific temperature and field strength). 
A representative sample of the data and fits at 100~K for a range of applied
longitudinal fields $B_{\rm app}$ is shown in Fig.~\ref{fig:100K_data}.
We plot the error against the applied field strength $B_{\rm app}$ in 
Fig.~\ref{fig:scaled_experimental_error_plot}, which is the experimental analogue of 
the plot in Fig.~\ref{fig:error_curve_fi}. 
At the lower temperatures, where the hopping is relatively slow, the results support 
the theoretical predictions made by our theory well: the standard deviation of the 
fitting parameter with the largest error (\(\nu\)) is minimal in the runs of the 
experiment where the field strength is the closest to the theoretical optimum 
for each temperature, and as a result, the error appears to be larger at field strengths 
that are not at the optimum. At higher temperatures, particularly in our case for 
temperatures above 200~K, the experimental data supports our theoretical predictions 
less convincingly, which is due to the muon hopping entering into the 
fast-fluctuation limit \cite{Hayano}, where the Kubo-Toyabe function takes for the form 
$G_{\rm DKT}\approx\exp(-\Delta^2 t/\nu)$. In these conditions, there is little information available to distinguish between $\nu$ values independently from $\Delta$ 
(in other words, there is a strong covariance between these parameters). 
As we are only concerned with $\nu$, $\Delta$ is a nuisance parameter and 
therefore there is little information about $\nu$ at these temperatures.

It is important to note that the method detailed here based on Fisher information 
calculates the \emph{minimum possible} error given an \emph{exact} fit function, 
but cannot take into account systematic errors (which leads to the error being slightly 
larger in reality), nor can it account for the fit function being an incomplete 
model of the data. In particular, including the muon-copper nuclear dipole coupling
and the copper quadrupolar terms in the muon's Hamiltonian explicitly leads to an 
additional spurious relaxation at late times \cite{Celio, Pratt}, leading to both 
an increase in the value of the apparent fluctuation rate and the associated 
error.

\section{Conclusion}


The method presented here is a fast and efficient method to 
calculate the lower bound on the 
experimental errors of the parameters of a given muon asymmetry function
as a function of the number of muon decay events. This method avoids 
computationally challenging data simulations, and provides an elegant
solution to working out the minimum number of muon decay events required to
obtain an error on the parameters of the muon asymmetry function. Application of
this to diffusion experiments suggests that the best
way of running such an experiment is to measure 50\% of the events at zero 
field, and the remaining events taken while applying just 
one optimally selected longitudinal field.

We note that the method presented here requires \emph{a priori} knowledge 
of the asymmetry function parameters before the measurement is taken, which 
is rarely 
possible. Practically, one approach to overcoming this could be to take short runs 
while cooling/warming the sample to get rough values of $\nu$ and $\Delta$ as a function 
of temperature, and then use Fig.~\ref{fig:heatmap_field} as a `rule of thumb' 
to obtain an estimate of the optimum field as a function of temperature.

\section{Code availability}
The code used to run the simulations and to generate the figures in this paper
is freely available at 
\url{https://github.com/ISISMuon/MuonExperimentOpt}. The scripts make use of the 
fit functions in the Mantid framework \cite{Mantid}, and can hence be adapted to 
other muon experiments in addition to those described here.

\printbibliography
\end{document}